\documentclass[aps,pre,floatfix,nofootinbib,10pt]{revtex4-2}
\usepackage{tabularx}
\usepackage{graphicx} \usepackage{amsmath} \usepackage{amssymb}
\usepackage[utf8]{inputenc}
\usepackage{xcolor}
\usepackage{lmodern}
\usepackage{multirow}
\newcommand{\BEA}{\begin{eqnarray}}
\newcommand{\EEA}{\end{eqnarray}}
\newcommand{\be}{\begin{eqnarray}}
\newcommand{\ee}{\end{eqnarray}}

\renewcommand{\varrho}{R}

\newcommand{\comment}[1]{}

\newcommand{\1}{{\cal P}_1}
\newcommand{\2}{{\cal P}_2}

\begin{document}

\title{Bargaining via Weber's law}

\author{V.G. Bardakhchyan$^{1,2)}$ \footnote{Corresponding 
author; e-mail: vardanbardakchyan@gmail.com
} and A.E. Allahverdyan$^{1,2)}$}

\affiliation{ 
$^{1)}$ Alikhanian National Laboratory (Yerevan Physics Institute), Alikhanian Brothers Street 2,  Yerevan 0036, Armenia,\\
$^{2)}$ Yerevan State University, 1 A. Manoogian Street, Yerevan 0025, Armenia 
}

\begin{abstract}
We solve the two-player bargaining problem employing Weber's law in psychophysics, which is applied to the perception of utility changes. Using this law, the players define the jointly acceptable range of utilities on the Pareto line, which narrows down the range of possible solutions. Choosing a unique solution can be achieved by applying the Weber approach iteratively. The solution is covariant to independent affine transformations of utilities. We provide a behavioral interpretation of this solution, where the players negotiate via Weber's law. For susceptible players, iterations are unnecessary, so they converge in one stage toward the (axiomatic) asymmetric Nash solution of the bargaining problem, where the weights of each player are expressed via their Weber constants. Thus the Nash solution is reached without external arbiters and without requiring the independence of irrelevant alternatives. We also show that our solution applies to the ultimatum game (which is not bargaining but still involves offer formation) and leads to an affine-covariant solution of this game that can reproduce its empirical features. Unlike previous solutions (e.g. the one based on fairness), ours does not involve comparing inter-personal utilities and is based on a partial symmetry between the proposer and respondent. 



\end{abstract}

\maketitle

\section{Introduction}
\label{intro}

\subsection{Theories of bargaining }

Axiomatic modeling of the two-player bargaining problem emerged from observations by von Neumann and Morgenstern on Pareto-optimal solutions and defection points \cite{luce}. Nash proposed a solution to the problem, based on 4 axioms \cite{luce,roth2012axiomatic}: {\it (i)} Covariance to affine utility transformations.  {\it (ii)} Symmetry between the two players. {\it (iii)} Pareto-optimality. {\it (iv)} Independence of irrelevant alternatives (sure-thing principle).

Several other axiomatic solutions have been proposed, such as Kalai-Smorodinsky, Kalai egalitarian, generalized Nash solutions (non-symmetric), and others \cite{peters2013axiomatic}. They generalize the four axioms in one way or another; e.g., the Kalai-Smorodinsky solution alters only the last axiom, known for its generally controversial status. The influence of bargaining theories is more general than behavioral economics, e.g. they find applications in thermodynamics, both conceptually \cite{babajanyan2020energy} and practically \cite{allahverdyan2023thermodynamic}. 

The drawback of axiomatic bargaining is that it does not provide a constructive framework for realistic negotiations. Zeuthen, and later Harsanyi, studied an iterative scheme of mutual compromises that led to the Nash solution \cite{luce,zeuthen,harsanyi1956approaches}. Rubinstein developed an iterative bargaining scheme with penalties for longer negotiations \cite{rubinstein1982perfect}, while Binmore {\it et al.} added to this model time-preferences \cite{binmore1986nash}. Both models revert to the Nash solution under certain conditions. Other iterative procedures were presented recently in Refs.~\cite{rachmilevitch2020implementation,karagozouglu2018iterated}. The first of them converges to the Nash solution. The rationale behind these proposals is sound, but (similar to Zeuthen's theory) they require an external arbiter who forces the players to do the required steps. Van Damme shows that when the players employ different concepts for the solution, they can negotiate sequentially, converging to the Nash solution under certain conditions that relate to risk aversion \cite{van1986nash}. The axiom {\it (iv)} is not required for this \cite{van1986nash}. Moulen shows how to implement sequentially the Kalai-Smorodinsky solution \cite{moulin1984implementing}. A set of lotteries is negotiated here, which is a more general framework than the one described below. Ref.~\cite{rubinstein1992interpretation} uses a similar framework to discuss the Nash axiomatic solution. Geometric interpretations of this solution are provided in Refs.~\cite{peters2013axiomatic,serrano1998beyond} \footnote{All these theories rely on the convexity of the bargaining set (i.e., the set of possible utility vectors). This convexity may be achieved via mixed strategies and expected utility-maximizing players \cite{luce}. In more general settings, allowing for non-convexity of the bargaining set goes against the symmetry axiom {\it (ii)} and leads to non-symmetric generalizations of the Nash solution \cite{zhou1997nash}. A similar result was obtained in Ref.~\cite{qin2015nash}, for a log-convex bargaining set.}.

\comment{
Van Damme in \cite{van1986nash} showed that under some conditions, when two players favor different methods, the Nash solution still persists, being dominant of all other strategies, in the following sense. If their solution does not lead to split, they continue bargaining, but each over new set of utilities, not exceeding the level of own utility required in previous step. Ultimately under these iterative scheme, they Nash solution happen to be optimal. The imposed requirement still forces the players to concede. In the paper the independence of irrelevant alternatives is relaxed, while other property is required, related to risk attitudes of players. It demands the change in risk attitude to be positively related to the outcome of the other player (the more risk averse is one player, the more the opponent gets).

For implementation of Kalai-Smorodinsky solution See paper by Moulen \cite{moulin1984implementing}.\footnote{{\color{blue} In the paper, however, agents bargain on the set of lotteries, which is to some extent more general framework than the one considered here. Within the same framework, Nash solution axiomatic foundation was discussed in \cite{rubinstein1992interpretation}.}} 
We confine ourselves to two-player convex (i.e. when the set of possible utility values is convex.) bargaining problem.}

\subsection{Weber's law}

It is clear that there are many theories devoted to axiomatic bargaining, but we believe it is necessary to develop a theory that appeals to the psychophysical and cognitive aspects of accepting (or not accepting) offers. Therefore, we develop a bargaining theory based on Weber's law, which is the foundation of psychophysics. A solvable example illustrating Weber's law is when a person already carries a certain (baseline) weight $S$ and should discriminate between $S$ and $S+\Delta S$ \cite{mather,akre}. 
More generally, the law is about discriminating two stimuli with intensities $S$ and $S+\Delta S$ given that $S$ refers to the baseline stimulus. In this context, baseline refers to a stimulus that arrived first and/or is more familiar to the discriminator. According to the law, those absolute differences $|\Delta S|$ between the two stimuli are discriminated for which 
\BEA
\label{wb}
|\Delta S|>kS, \quad k>0,
\EEA
where $k>0$ is a constant that does not depend on $\Delta S$ and $S$ but depends on the nature of the stimulus and the discriminator (individual). The minimal absolute difference $|\Delta S|=kS$ defines the concept of the just-noticeable-difference \footnote{Note the difference between Weber's and the the Weber-Fechner law. One possible derivation of the Weber-Fechner law is as follows.
One assumes equality between $|\Delta S|/(kS)$ and the subjective sensation $\Delta \psi$, which is a new quantity. Then integrating this relation leads to $\psi=k\ln S+a$, where $a$ is an integration constant \cite{mather}. Our focus will not be on the Weber-Fechner law here. But we emphasize that the validity of the Weber-Fechner law supports Weber's law itself.   
Note that the Weber-Fechner law is better studied in consumer behavior than the proper Weber law \cite{monroe1973buyers, white2014psychophysics}.}. Limitations of the Weber law are discussed in \cite{mather,luce2} \footnote{A serious limitation that we omit in the present work is that Weber's law is probabilistic and in its generalized form refers to the probability of discrimination; see \cite{kacelnik1998risky} for a review of this aspect. A generalization of Weber's law is that the probability of discrimination is proportional to $\Delta S/S$. \\ One can try to interpret (\ref{wb}) in terms of internal variability or fluctuations, where $|\Delta S|$ is the characteristic magnitude of fluctuations. Then (\ref{wb}) refers to a specific (e.g. Gaussian) random variable, where $S$ is the mean signal. }. Note that (\ref{wb}) does not depend on the units chosen to measure $S$. This scale-invariance feature will be crucial for applications of (\ref{wb}) to bargaining. 

Weber's law has traditionally been applied to the perception of weight, loudness, and brightness, i.e., phenomena where stimulus compression occurs at the peripheral level \cite{mather,akre}. This is the traditional domain of psychophysics. More recently, the law was observed and validated on systems that are ``cognitive'' rather than ``sensory'', e.g., the perception of temporal durations \cite{wearden}, or cognitive representation of number in humans and animals \cite{dehaene1998abstract,dehaene}. 
These findings are supported both by behavioral and neurophysiological evidence \cite{dehaene}.
The stimuli here can be an abstract concept, e.g. Arabic numbers. 

In 1956, Luce applied Weber's law to his consumer choice model \cite{luce_jnd}; see \cite{gilboa2009theory} for a review. Recently, Argenziano and Gilboa related the Weber law to the Cobb-Douglas utility \cite{gilboa_cobb}, and applied the law to the concept of interpersonal utility comparison \cite{gilboa_interpersonal_utility}. Weber's law was also applied in price perception \cite{monroe1973buyers}, though such applications require caution \cite{kamen1970psychophysics}; see Ref.~\cite{white2014psychophysics} for a recent review. 
The idea was to take the price of a product as $S$ and analyze its price change $|\Delta S|$ which the consumer does not discriminate. In the context of money or wealth, it is more reasonable to speak about tolerance to a change rather than not perceiving it. Weber's law applies to other aspects of consumer behavior (product design, packaging, size of advertisements, frequency of presentation, {\it etc.}) \cite{britt1975weber}.

\subsection{Our approach}

Our approach starts with an observation that people typically negotiate about sufficiently large sums of money in larger amounts. For example, people may not focus on differences in house price that are comparable to their monthly salary. 
Hence, in such negotiations, one can distinguish between the baseline proposal for the price and an acceptable domain of prices around it. Taking this logic to a two-player bargain on the Pareto line, we propose (for the first player) a bargaining strategy that specifies a baseline utility and an acceptable utility domain defined via Weber's law. If the second player follows the same strategy, we can postulate from symmetry that the acceptable domains for both players coincide. The end-points of the domain are the baseline utilities for each player. This emergent concept of the joint acceptable domain already restricts the initial freedom of choices that amounts to the whole Pareto line (restricted by defection points). Alternatively, the joint acceptable domain can be reached behaviorally, through a sequence of mutual proposals that respect Weber's law.  

There is an important particular of susceptible players (both $k_1$ and $k_2$ are small), where the joint acceptable domain asymptotically squeezes toward a single point. This point turns out to be the asymmetric, axiomatic Nash solution with the weights given by $k_2/(k_1+k_2)$ and $k_1/(k_1+k_2)$ for (resp.) first and second player; see (\ref{kon}) below. Hence the Nash solution is reached without postulating the sure-thing principle (see \cite{we_regret} for a recent criticism of this principle), and without external arbiters. The symmetric axiomatic Nash solution is recovered for $k_1=k_2$, i.e., equal susceptibilities of the players.

More generally, when the joint acceptable domain does not tend to a single point, we propose another axiom that allows the players to reach a preliminary compromise and share the bargaining resource accordingly. Now the bargaining can be continued over the remaining resource, and if it follows the same strategy as above then after sufficiently many steps of repeating the above process the players can asymptotically reach a final agreement. 

We applied Weber's law to the ultimatum game \cite{guthfairisunfair,fehr,ultimatum}. This asymmetric (i.e. the players have different roles) game emerged as a basic challenge to the game-theoretic rationality, because people here do not behave according to the Nash equilibrium from the non-cooperative game theory. The ultimatum game does not involve proper bargaining, but it still includes a non-trivial offer made by the first player. We argue that this offer can be generated via Weber's law using the above concept of the joint acceptable domain. Once only the first player makes an offer, it is reasonable to assume this player offers according to her baseline utility. We show that the resulting theory is capable of describing several non-trivial experimental features of the ultimatum game. As compared to existing solutions of the ultimatum game [see \cite{ultimatum} for a recent review], the proposed solution has two pertinent advantages. First, it does not involve inter-personal utility comparison (as e.g., the fairness theory does). Second, it can generate predictions for a wider range of behavioral strategies for the proposer and respondent; see section \ref{ulti}.

\subsection{Organization of the paper}

In the next section, we recall the standard bargaining setup. Unlike standard presentations of such setups (see e.g. \cite{roth2012axiomatic}), we emphasize the resource over which bargaining occurs. Section \ref{axioms} applies Weber's law to this setup. Several axioms are proposed that provide a solution to the bargaining problem. Section \ref{behave} shows that the main result of the axiomatic approach can be obtained via a behavioral approach of sequential proposals. A simple example illustrating our approach is also provided in this section. Section \ref{ito} explores iterative bargaining, which allows to converge to a unique solution. Our results for the ultimatum game, which includes proposer and responder, are shown in section \ref{ulti}. This game is not proper bargaining but still involves nontrivial offer by the proposer that takes into partial account the interests of the responder. The last section summarizes our results and suggests several open problems.

\section{Bargaining set-up }

We study two-player bargaining and make the following standard assumptions \cite{roth2012axiomatic}. $X>0$ is the amount of a resource, which the two players $\1$ and $\2$ bargain to split. For simplicity, we assume that this resource is infinitely divisible. Our results generalize to the case in which only finite values of $X$ are available for the distribution between the players; we do not address this case. 

Let $x_1\geq 0$ and $x_2\geq 0$ be the amount of resource got by (resp.) $\1$ and $\2$. The utility functions of $\1$ and $\2$ are (resp.) 
\BEA
\label{baradar}
U_1 (x_1), \qquad U_2 (x_2).
\EEA
We assume that $U_1(x)$ and $U_2(x)$ are strictly increasing, sufficiently smooth, and concave \cite{luce}: 
\BEA
\label{taliban}
U_k' (x)>0, \qquad U_k'' (x)\leq 0, \qquad k=1,2.
\EEA
The concavity assumption means that as the obtained resource increases, the additional utility gained from each additional unit decreases (the principle of diminishing marginal utility) \cite{luce}. 

On the Pareto line, the available resource is fully allocated between the players, leaving no surplus or waste.
Thus, in the utility space, the Pareto line is defined via a strictly decreasing, smooth, concave function $h[\cdot]$:
\BEA
\label{h-function}
  &&  u_2=h[u_1],\quad h'[y]<0,\quad h''[y]\leq 0,\\
  && h[u_1]\equiv U_2(X-U_1^{-1}(u_1)),
    \label{bo}
\EEA
where $h'[y]<0$ and $h''[y]\leq 0$ in (\ref{h-function}) follow from differentiating (\ref{bo}) and using (\ref{taliban}).
The same features hold for the inverse function 
\BEA
\label{g-function}
u_1=g[u_2],\quad g\Big[h[u]\,\Big]=u,
\quad g'[y]<0,\quad g''[y]\leq 0.
\EEA
If (\ref{h-function}) is taken alone (without mentioning the resource), then the Pareto line is presented in terms of utility, and the existence of the underlying resource is hidden. This refers to a welfarist approach \cite{roth2012axiomatic}. For our purposes, working with welfarist approaches is insufficient; see e.g. section \ref{ito}. 

If no agreement is reached in the bargaining process, the players receive (defection) utilities $d_1$ and $d_2$. Naturally, we demand: 
\BEA
\label{mo}
U_1(x_1)\geq d_1, \qquad U_2(X-x_1)\geq d_2, 
\EEA
for all allowed $x_1\geq 0$ and $X-x_1\geq 0$. 

\section{Axioms of Bargaining }
\label{axioms}

{\it Axiom 1.} $\1$ and $\2$ operate on the Pareto line (\ref{h-function}). Hence a larger $u_1$ means a smaller $u_2$, which is the main conflict to be solved via bargaining. 

{\it Axiom 2.} $\1$ and $\2$ hold Weber's law for utility changes (\ref{wb}) with (resp.) coefficients
\BEA
1>k_1>0,\qquad 1>k_2>0.
\label{zebra}
\EEA
Now holding Weber's law for $\1$ means that $\1$ has a baseline utility $ u_1$, and the acceptable range of utilities is defined as 
\BEA
\label{koala}
\frac{ u_1- u_1^*}{ u_1-d_1}=k_1,
\EEA
where $ u_1$ is naturally compared with the defection utility $d_1$, where $u_1^*$ defines the border of acceptance given $ u_1$. 

\begin{figure}[!ht]
\centering    \includegraphics[scale=0.8]{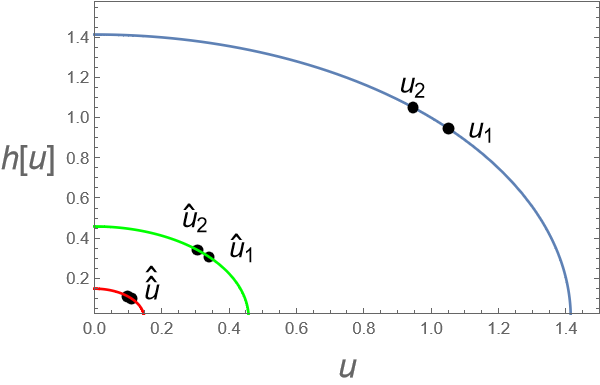}
\caption{The function $h[u]=\sqrt{X-u^2}$ for various values of the resource $X$ and $p=q=0.5$; see (\ref{h-function}, \ref{su11}, \ref{su12}). The Weber coefficients are $k_1=k_2=0.1$; cf.~(\ref{wb}, \ref{zebra}). The defection point is $(0,0)$. Blue line: $h[u]$ for $X=2$. Now $u_1$ and $u_2$ are solutions of (\ref{simba}--\ref{simba3}) (black dots on the blue line).
Initial set of solutions included the whole Pareto line (blue line). After applying {\it Axioms 1-3}, the solution is restricted to the part of this curve between $u_1$ and $u_2$ (black points on the blue line).\\
The first partial sharing of the resource brings players from $X=2$ to $X=0.21$; see section \ref{ito}. The green line shows $h[u,X=0.21]$. The corresponding solutions of (\ref{simba}--\ref{simba3}) are denoted by $\hat{u}_1$ and $\hat{u}_2$ (black dots on the green line). It is seen that $\hat{u}_1$ and $\hat{u}_2$ are closer to each other than $u_1$ and $u_2$. The second partial sharing of the resource leads from $X=0.21$ to $X=0.022$. The red line shows $h[u,X=0.022]$. Now the solutions of (\ref{simba}--\ref{simba3}) nearly coincide, i.e., $\hat{\hat{u}}=\hat{\hat{u}}_1\approx \hat{\hat{u}}_2$ (see black dots on the red line).
}
\label{figure}
\end{figure}

{\bf Remark 1.} Note that (\ref{koala}) is invariant under affine transformations of utilities: 
\BEA
\label{affo}
 u_1\to a_1 u_1+b_1, \quad u_1^*\to a_1u_1^*+b_1, \quad d_1\to a_1 d_1+b_1, \qquad a_1>0.
\EEA
Such transformations do not change the utility content \cite{luce}. The solution of the bargaining problem will be based on (\ref{affo}), and hence it will inherit its behavior under affine transformations. Thus, the affine-covariance of the bargaining solution need not be imposed as an additional axiom. Below we assume $d_1=d_2=0$ for simplicity. This is allowed due to (\ref{affo}). Now the allowed domain of utilities is $\{u_1\geq 0, u_2\geq 0\}$ limited by a convex curve; 
see Fig.~\ref{figure}.

The analog of (\ref{koala}) can be written for $\2$ ($d_2=0$):
\BEA
\label{koala2}
\frac{ u_2- u_2^*}{ u_2}=k_2,
\EEA
with the baseline utility $u_2$ and the border utility $u_2^*$.

{\it Axiom 3} states the permutation symmetry between $\1$ and $\2$ in the following sense: acceptable domains (\ref{koala}) and (\ref{koala2}) coincide with each other: $u_1^*= u_2$ and $u_2^*= u_1$. Hence, (\ref{koala}) and (\ref{koala2}) can be written as [see (\ref{g-function})]
\begin{align}
\label{simba}
&\frac{u_1-g[u_2]}{u_1}=k_1, \qquad \frac{u_2-h[u_1]}{u_2}=k_2,
\end{align}
or equivalently as
\begin{align}
\label{simba2}
&u_1=E[u_1], \qquad
E[u_1]\equiv g\Big[(1-k_2)h[(1-k_1)u_1]\,\Big], \\
&u_2=F[u_2], \qquad
F[u_2]\equiv h\Big[(1-k_1)g[(1-k_2)u_2]\,\Big].
\label{simba3}
\end{align}
Eqs.~(\ref{simba2}) define two different equations for two unknowns $u_1$ and $u_2$. The solutions are unique, as the following proposition shows. Note that the symmetry in the sense of {\it Axiom 3} is compatible with different utilities (for $\1$ and $\2$), and with $k_1\not=k_2$. The concept of symmetry can be used also in a different sense, i.e., in the sense of equal utilities or $k_1=k_2$; see section \ref{ulti}.

{\bf Proposition 1:} Solutions $u_1$ and $u_2$ of (\ref{simba2}, \ref{simba3}) are unique. This is shown from (\ref{h-function}, \ref{g-function}, \ref{zebra}) by noting that $E'[u_1]>0$
and $F'[u_2]>0$. $\square$

{\bf Remark 2.} Note that (\ref{simba}--\ref{simba3}) do not specify a unique solution. But they still do restrict the initial solution range: initially we had the whole Pareto line. Now solutions are limited by $u_1$ and $u_2$; see Fig.~\ref{figure}.
In section \ref{behave}, we shall interpret $u_1$ and $u_2$ from (\ref{simba}--\ref{simba3}) as two confronting offers made by $\1$ and $\2$, respectively. In section \ref{ito}, we discuss how to implement (\ref{simba}) iteratively to reach a unique solution. 

There is however an important particular case, where iterations are not necessary and (\ref{simba}--\ref{simba3}) provide asymptotically a unique solution.

{\bf Proposition 2.} Consider susceptible players: $k_1\to 0$ and $k_2\to 0$, but $k_1>0$, $k_2>0$, and $k_1/k_2={\cal O}(1)$. We now get from (\ref{simba2}, \ref{simba3}):
\BEA
\label{kon}
&& u_1=h[u_2]+{\cal O}(k_1),\\
&& u_1=u_1^*+{\cal O}(k_1), \quad u_1^*
\equiv {\rm argmax}_{u}\mu[u],\quad \mu[u]\equiv u^{\frac{k_2}{k_1+k_2}} h[u]^{\frac{k_1}{k_1+k_2}}.
\label{akon}
\EEA
The maximization in (\ref{akon}) can be performed by calculating $\mu'[u]=0$ and checking that $\mu''[u]<0$ due to (\ref{h-function}, \ref{zebra}). Now by Taylor-expanding (\ref{simba2}) for $k_1\to 0$ and $k_2\to 0$, and keeping quantities ${\cal O}(k_1)$ we obtain: $k_1u_1h'[u_1]=-k_2h[u_1]$, which is the same as $\mu'[u_1]=0$. Taylor-expanding (\ref{simba3}) produces $u_2=g[u_1]+{\cal O}(k_1)$ and leads to (\ref{kon}). $\square$

{\bf Remark 3.}
Note that $u^{\frac{k_2}{k_1+k_2}} h(u)^{\frac{k_1}{k_1+k_2}}=u_1^{\frac{k_2}{k_1+k_2}} u_2^{\frac{k_1}{k_1+k_2}}$ in (\ref{akon}) is precisely the weighted product of utilities that appears in the (asymmetric) Nash solution \cite{roth2012axiomatic}. Here $\frac{k_2}{k_1+k_2}$ and $\frac{k_1}{k_1+k_2}$ are the weights for $\1$ and $\2$, respectively. Due to $k_1/k_2={\cal O}(1)$ both weights can be finite despite $k_1\to 0$ and $k_2\to 0$. Thus, the axiomatic Nash solution can have a psychophysical interpretation. Moreover, it also provides a sound way to attribute weights in the asymmetric Nash solution. This solution is meaningless without a concrete weight-attributing mechanism; for arbitrary weights, the solution covers the whole Pareto line. As expected, a more susceptible player (i.e., the first player if $k_1<k_2$) has a larger weight in (\ref{akon}). However, for different utilities this does not mean that a more susceptible player necessarily demands more resources; see (\ref{afgan11}).

\section{Behavioral interpretation}
\label{behave}

Solution (\ref{simba}) can be given a behavioral interpretation via three rules. The second rule is based on Weber's law. 

{\it Rule 1.} $\1$ and $\2$ exchange proposals about the proper solution. Proposals are made on the Pareto line. 

{\it Rule 2.} Let $\1$ proposes a solution with utility $u^{[1]}_1$. Following Weber's law, $\2$ takes this proposal as the baseline of $\1$, and counter-proposes the best option for $\2$, which is still in the acceptance range of $\2$. I.e., $\2$ proposes $u^{[1]}_2$, where $(u^{[1]}_1-g[u^{[1]}_2])/{u^{[1]}_1}=k_1$; cf.~(\ref{simba}). Likewise, $\1$ takes the proposal $u^{[1]}_2$ by $\2$ as the baseline by $\2$ and proposes $u_1^{[2]}$ defined from $(u_2^{[1]}-h[u_1^{[2]}])/{u_2^{[1]}}=k_2$.
Altogether, these (counter)proposals define two recursions [cf.~(\ref{simba2}, \ref{simba3})]:
\begin{align}
\label{simba4}
&u^{[k+1]}_1=E[u_1^{[k]}], \\
&u_2^{[k+1]}=F[u_2^{[k]}], 
\label{simba5}
\end{align}

{\it Rule 3.} Above proposals continue till no novel proposals are generated along {\it Rule 1} and {\it Rule 2}. 

Note that there is no meaning for (say) $\2$ in accepting $u^{[1]}_1$, because $\1$ will take this agreement as the baseline for $\2$ and will then generate a proposal that is worse for $\2$ than $u^{[1]}_1$. In this sense, {\it Rule 3} is sensible, since proposals do not have an external reinforcement. Such a reinforcement from one side will violate the symmetry between $\1$ and $\2$. {\it Rule 3} is supported by the following proposition.

{\bf Proposition 3.} The sequences (\ref{simba4}) and (\ref{simba5}) converge to (\ref{simba}--\ref{simba3}). We shall prove the convergence of (\ref{simba4}) only. First we show that $0<E'[u]<1$:
\BEA
E'[u]=(1-k_1)(1-k_2)g'[(1-k_2)h[(1-k_1)u]]h'[(1-k_1)u]=
(1-k_1)(1-k_2)\frac{g'[(1-k_2)h[(1-k_1)u]]}{g'[h(1-k_1)u]]},
\label{gog}
\EEA
where in the last line we used $g'[h[x]]h'[x]=1$ that follows from $g[h[x]]=1$; cf.~(\ref{g-function}). Now in (\ref{gog}) we employ $g''[x]\leq 0$ and $g'[x]<0$ to deduce that the fraction in (\ref{gog}) is $\leq 1$. The desired $0<E'[u]<1$ follows from (\ref{zebra}). Let us now employ the mean-value theorem
\BEA
\Big |E[u_1^{[k+1]}] - E[u_1^{[k]}] \Big |=\Big |u_1^{[k+1]} - u_1^{[k]} \Big |\, E'[c]
< \Big |u_1^{[k+1]} - u_1^{[k]} \Big |,
\label{soroka}
\EEA
where $c\in(u_1^{[k+1]}, u_1^{[k]})$, and where in the last line of (\ref{soroka}) we used $0<E'[c]<1$. The last inequality in (\ref{soroka}) shows that the sequence generated from (\ref{simba4}) converges to a fixed point, which is unique according to {\bf Proposition 1}. $\square$

{\bf Example.}
Take for example in (\ref{baradar}): 
\BEA
\label{su11}
& U_1(x_1)=x_1^{p},~~ U_2(x_2) =x_2^{q} = (X-x_1)^{q}, \\
& u_2 = h(u_1)=\left(X-u_1^{1/p}\right)^{q},~~ p,q<1.
\label{su12}
\EEA
Ref.~\cite{galanter} presents empirical support for utility functions (\ref{su11}) and estimates parameters $p$ and $q$. 

Eqs.~(\ref{simba2}, \ref{simba3}) have the following solutions [cf.~(\ref{zebra})]:
\BEA
\label{afgan1}
&& x_1= (u_1)^{1/p} = \epsilon_1 X, \quad x_2=(u_2)^{1/q} = \epsilon_2 X, \\
&& \label{afgan2}  
\epsilon_1= 
\frac{1-(1-k_2)^{1/q}}{1-(1-k_1)^{1/p}(1-k_2)^{1/q}}, \quad
\epsilon_2 = 
\frac{1-(1-k_1)^{1/p}}{1-(1-k_2)^{1/q}(1-k_1)^{1/p}},\quad 0<\epsilon_1,\epsilon_2<1,
\EEA
where $x_1$ and $x_2$ are the resources that refer to $u_1$ and $u_2$, respectively. 
According to (\ref{afgan1}, \ref{afgan2}), the amount of resource demanded by each player depends on the utilities and $k$-coefficients of both players. Note that $x_1+x_2\geq X$, as it should be for conflicting demands. 
The approximate equality $x_1+x_2\approx X$ is reached, when $k_1$ and $k_2$ in (\ref{afgan1}) are sufficiently small [cf.~{\bf Proposition 2}]:
\BEA
\label{afgan11}
\epsilon_1\simeq\frac{(k_2/q)}{(k_2/q)+(k_1/p)}, \quad 
\epsilon_2\simeq \frac{(k_1/p)}{(k_2/q)+(k_1/p)}. 
\EEA

\section{Iterative bargaining}
\label{ito}

In view of {\it Rules 1-3}, Eq.~(\ref{simba}) means that $\1$ and $\2$ are involved in the deadlock of making contradicting proposals $u_1$ and $u_2$ (resp.) that hold Weber's law for each one. To get out of this deadlock we propose the following.

{\it Axiom 4.} Given (\ref{simba}), $\1$ and $\2$ agree on a compromise: they temporarily go out of the Pareto line, and $\1$ [$\2$] takes the amount of the resource proposed by $\2$ [$\1$]. In other words, $\1$ [$\2$] takes the resource that refers to the utility $h[u_2]$ [$g[u_1]$]. After this step, they are back to bargaining over the remaining resource. 

If the players $\1$ and $\2$ apply the same {\it Axioms 1-3}, then the analogues of (\ref{simba}) will be reached over the new resource. Such iterations will eventually converge, since the amount of the resource keeps on decreasing from one iteration to another; see Fig.~\ref{figure}. 

Let us illustrate this iteration process via the above example (\ref{su11}, \ref{su12}).
The resources obtained (resp.) by $\1$ and $\2$ read:
\BEA \label{epsilons}
(1-\epsilon_2)X\sum_{m=0}^\infty(\epsilon_1+\epsilon_2-1)^m=\frac{(1-\epsilon_2)X}{2-\epsilon_1-\epsilon_2}, 
\qquad
(1-\epsilon_1)X\sum_{m=0}^\infty(\epsilon_1+\epsilon_2-1)^m=\frac{(1-\epsilon_1)X}{2-\epsilon_1-\epsilon_2}, 
\EEA
Hence the resource is now fully shared, the corresponding weights being $\frac{1-\epsilon_2}{2-\epsilon_1-\epsilon_2}$ and $\frac{1-\epsilon_1}{2-\epsilon_1-\epsilon_2}$. For $k_1=k_2$ and $p=q$, we find that the resource is shared equally due to $\epsilon_1=\epsilon_2$; see (\ref{afgan2}).

Note that iteratively converged solutions (\ref{epsilons}) do not coincide with the Nash axiomatic solution when $k_1=k_2$ is not small, but $p\not=q$. It is also important to remember that we assumed no changes in Weber constants between iterations.

\comment{
{\color{blue} For other utilities where symmetry between player's in sensitivity (k-s) are same but their utilities may still differ, it can be shown that $1>k_1=k_2>>0$ does not entail the same conclusion for Nash non-symmetric and iterative bargaining. 
In \ref{epsilons}, taking $p\neq q$ verifies that $\frac{\epsilon_1}{\epsilon_1+\epsilon_2}$ does provide other split than optimizing \ref{ko}. As a numerical example, with $X=1; p=0.5$ and $q=0.25$ the iterative bargaining split is $0.7:0.3$ when $k_1=k_2=0.235$, while Nash non-symmetric solution yields $0.781:0.219$ split.}
}

\section{Ultimatum game}
\label{ulti}

\subsection{Definition of the ultimatum game}

Here the first player $\1$ makes a proposal about dividing a fixed resource $X$ (e.g., a fixed sum of money), and the second player $\2$ may either agree with the proposal or reject it, which reverts the players to defection points (zero resource for both) \cite{guthfairisunfair,fehr,ultimatum}. Hence the situation is essentially asymmetric and does not involve real bargaining. The formal solution of the game proceeds via the unique subgame perfect Nash equilibrium, where the first player offers the smallest possible (but strictly positive) amount of resource. Indeed, for the second player it is rational to accept any positive offer. Knowing this, the first player proposes the smallest positive amount. The same result is obtained within Stackelberg's equilibrium. 

However, in practice people normally violate this subgame perfect equilibrium solution, because the first player offers the second between $50\%$ and $\sim 75\%$ of the resource $X$ \cite{guthfairisunfair,fehr,ultimatum}. Several other experimental features of the ultimatum game are reviewed below. 

Several approaches were proposed for explaining this seemingly irrational behavior; see \cite{ultimatum} for a recent review of various approaches devoted to the ultimatum game. The most known of them is based on fairness, which (for equal utilities of the players) states that the proposer chooses somewhere between the equal split $X/2$ and the Nash equilibrium \cite{guthfairisunfair,fehr}. An alternative explanation is based on the principle {\it "It hurts me more than it hurts you"} and states that the responder rejects offers, whenever the responder loses less than the proposer \cite{ultimatum}. These approaches are different, although they are capable of describing existing experiments. Their joint drawback is that they involve an inter-personal utility comparison [i.e., these solutions are not affine-covariant in the sense of (\ref{affo})], a condition that prevents taking any of them as a normative solution for the ultimatum game \footnote{Note that for equal utilities of the players the fair solution $X/2$ coincides with the axiomatic Nash solution for the bargaining. However, for inequal utilities, the fair solution is just opposite to the affine-covariant Nash axiomatic solution. This point is discussed in detail in Ref.~\cite{ultimatum}.}. 

\subsection{Solution of the ultimatum game based on Weber's law}

We recall again that the ultimatum game is not proper bargaining, i.e., it is a one-shot game. However, it does involve the offer formation from $\1$ (the proposer) that should account for a possible rejection from $\2$ (responder). 

We start from (\ref{simba}). Now $\1$ offers $\2$ utility $u_1$ for herself and hence utility $g[u_1]$ for $\2$. The interests of $\2$ are accounted for because $g[u_1]$ is in the boundary of the acceptance domain of $\2$. This is our scenario of applying Weber's law to the ultimatum game. 

To explore the ultimatum game in this scenario, we shall focus on the well-known logarithmic utility functions that emerged in the early days of decision theory \cite{luce}. These utilities depend on the previous resources $\gamma_i$ (e.g. money) of the players and read for (resp.) the first and second player \cite{luce}: $\ln (\gamma_1+x_1)$ and $\ln (\gamma_2+x_2)$, where $d_1=\ln (\gamma_1)$ and $d_2=\ln (\gamma_2)$. Recalling that we work with $d_1=0$ and $d_2=0$, the utilities after the suitable affine-transformations (\ref{affo}) read:
\BEA
u_i(x_i)=\ln(1+\frac{x_i}{\gamma_i}),\qquad i=1,2.
\label{orto}
\EEA
The total resource is still denoted $X$. Now (\ref{simba2}) and (\ref{orto}) are solved numerically.

\subsection{The players with equal utilities and $k_1=k_2$}

\begin{table*}
\caption{Ultimatum game with $\gamma_1=\gamma_2$ and $k_1=k_2$. The first player is the proposer.
For various values of the parameters, we show the share ${x}_1$ of the first player, which is always larger than $X/2$. 
Here $x_1$ is the resource demanded by $\1$ that refers to utility $u_1$. 
}
\begin{tabularx}{0.99\textwidth} { 
  || >{\centering\arraybackslash}X 
  | >{\centering\arraybackslash}X   | >{\centering\arraybackslash}X | >{\centering\arraybackslash}X |>{\centering\arraybackslash}X |
    }
 \hline
 $(k_1=k_2; ~\gamma_1=\gamma_2; ~X)$ & $( 0.01; 5; 10^3  )$ & $( 0.01; 500; 10^3 )$  & $(0.1 ; 5; 10^3  )$ & $(0.1 ; 5; 10^2)$ \\
 \hline \hline
 ${x}_1$  & $511.7$  & $503.5$  & $619.6$ & $56.9$ \\
\hline
\end{tabularx}
\label{tab1}
\end{table*}

We start with the situation, where the parameters of both players are identical, $\gamma_1=\gamma_2$, $k_1=k_2$, and the only asymmetry between them relates to $\1$ being the proposer. 
Using (\ref{simba2}, \ref{orto}), we draw the following conclusions, which are illustrated in Table~\ref{tab1}. 

{\it (i)} Let $x_1$ be the resource demanded by $\1$ that refers to utility $u_1$. 
The first player (proposer) gets more than $X/2$: ${x}_1>X/2$. This effect is widely seen in experiments \cite{guthfairisunfair,fehr,stakes,ultimatum}. In our situation, this is a consequence of the first-choice right, which is (for the ultimatum game) not confronted by the analogous offer from the second player, i.e. we look at (\ref{simba2}) only.

\comment{We emphasize that $\widetilde{x}_1>X/2$ cannot be explained via the Nash (axiomatic) solution which predicts $\widetilde{x}_1=X/2$. Likewise, the (non-cooperative) Nash equilibrium cannot explain any other solution $X/2 <\widetilde{x}_1<X$.}

{\it (ii)} For more susceptible players (i.e. smaller $k_1=k_2$) $\frac{{x}_1}{X/2}$ is closer to $1$. 

{\it (iii)} For richer agents (i.e. larger $\gamma_1=\gamma_2$) $\frac{{x}_1}{X/2}$ is closer to $1$. 
Hence, two richer agents will reach a fairer agreement than two poor ones. These results agree with experimental findings reported in Ref.~\cite{boarini2009interpersonal}.

{\it (iv)} For a smaller $X$, $\frac{{x}_1}{X/2}$ is closer to $1$; i.e., a smaller amount of resources is divided more fairly. This prediction of our approach coincides with experiments \cite{stakes,ultimatum,andersen2011stakes}. The finding was explained via a cost of fairness: for larger stakes people do not want fairness, they want money, i.e. tend to accept relatively small, but now absolutely large offers \cite{stakes}. Our explanation of this effect emphasizes not the fairness cost of the second (acceptor) player, but rather the susceptibility of the players. 

\subsection{The players with inequal utilities and $k_1\not=k_2$}

\begin{table*}
\caption{{Ultimatum game with asymmetric players: $\gamma_1\not=\gamma_2$ and $k_1\not=k_2$.
For various values of the parameters, we show the share ${x}_1$ of the first player. }
}
 { \begin{tabularx}{0.99\textwidth} { 
  || >{\centering\arraybackslash}X 
  | >{\centering\arraybackslash}X   | >{\centering\arraybackslash}X | >{\centering\arraybackslash}X |>{\centering\arraybackslash}X |
    }
 \hline
 $(k_1;k_2; ~\gamma_1; \gamma_2; ~X)$ & $( 0.1; 0.01; 5; 5; 10^3  )$ & $( 0.01; 0.1; 5; 5; 10^3 )$  & $(0.01 ; 0.01; 50; 5; 10^3  )$ & $(0.01 ; 0.01 ; 5; 50; 10^3)$ \\
 \hline \hline
 ${x}_1$  & $151.2$  & $894.2$  & $621.1$ & $396.1$ \\
\hline
\end{tabularx} }
\label{tab2}
\end{table*}

To our knowledge, no ultimatum-game experiments so far checked systematically how players with different backgrounds make proposals. Here are predictions of our solution; see Table~\ref{tab2}. 

{\it (v)} For different coefficients $k_1$ and $k_2$ in (\ref{zebra}), a more susceptible player tends to get more. In other words, ${x}_1$ is a decreasing function of $k_1$, and an increasing function of $k_2$. For example, the value $511.7$ in Table~\ref{tab1} becomes $151.2$ for $(k_1=0.1; k_2=0.01)$, and $894.2$ for $(k_1=0.01; k_2=0.1)$; see Table~\ref{tab2}. Note that ${x}_1>X/2$ no longer holds for asymmetric players, i.e. the proposer need not take more than one-half of the resource. 
One interpretation is that a more susceptible respondent is proposed more, because such a respondent is more likely to reject.

{\it (vi)} Different initial resources: the rich tend to become even richer. Now ${x}_1$ is a decreasing function of $\gamma_2$, and an increasing function of $\gamma_1$. For example, the value $511.7$ in Table~\ref{tab1} becomes 
$621.1$ for $(\gamma_1=50; \gamma_2=5)$, and $396.1$ for $(\gamma_1=5; \gamma_2=50)$.

Some direct conclusions to be drawn from {\it (v)} and {\it (vi)} are as follows: a richer proposer can offer less than $X/2$, but only when he/she is less susceptible towards giving up money (philanthropist). A rich and susceptible proposer would like to take almost the whole $X$ (greedy person). Only such greedy cases effectively hold the sub-game perfect Nash equilibrium. 

The Weber constant is generally dependent on the wealth of the individual, and this dependence was not taken into account above.

\section{Summary}

We proposed a solution for bargaining that employs Weber's law, a basic law in psychophysics that has been known for two centuries and limits the sensory and cognitive abilities of human decision-making \cite{mather,akre,thurstone1927three}. 
The solution starts by stating the concepts of (ir)relevance and the just noticeable differences implied by Weber's law. Next, each player defines the baseline utility and the domain of acceptable utilities on the Pareto line. There are two equivalent ways to make these domains consistent with each other. The first way is to use the symmetry axiom, which postulates that the domains coincide. The second, behavioral way assumes that the players make proposals within the acceptable domain employing the previous proposal of the opponent as her baseline. Our implementation of Weber's law reduces the set of allowed solutions, but generically does not produce a unique solution, unless new axioms are applied. An important exclusion from this is the particular case of susceptible (in Weber's sense, i.e., the just-noticeable differences are small) players, where our solution leads to the asymmetric Nash solution with weights expressed via the Weber coefficients of the players. Hence, our results show the Nash solution can have psychophysical foundations. More generally, the Weber strategy of bargaining can be implemented iteratively via an additional natural {\it Axiom 4}, and it converges to a unique solution, where the resource is exhausted. 

Altogether, our solution employs four axioms. {\it Axiom 1} (Pareto-optimality) and {\it Axiom 3} (symmetry) find their analogs in axioms that led to the Nash solution; see the beginning of section \ref{intro}. As compared to those axioms, we do not employ the axiom on affine-covariance, because this covariance follows from Weber's law, which is our {\it Axiom 2}. We also do not employ the sure thing principle; see axiom {\it (iv)} in section \ref{intro}. Our {\it Axiom 3} can be replaced with three rules formalizing sequential bargaining that satisfies Weber's law. Appendix \ref{appendix} presents an alternative application of Weber's law to bargaining. Although this application is less transparent than {\it Axiom 2} and {\it Rules 1-3}, it utilizes somewhat different concepts for conflict and compromise in the bargaining process. 

We show that our results apply to the ultimatum game, which does not involve real bargaining, but still closely relates to offer formation. The challenge of the ultimatum game is that people do not follow its sub-game perfect Nash equilibrium, thereby defying the basic tenet of classical rationality. Our approach to solving the ultimatum game assumes that the proposer acknowledges the interests of the responder by making the offer right on the verge of respondent's acceptance domain. Our solution of the ultimatum game is affine-covariant, i.e., it does not involve inter-person utility comparison. This contrasts previous solutions that involved such comparisons; see \cite{ultimatum} for a recent review. The solution depends on the initial wealth of the proposer and respondent, as well as on their Weber constants. It reproduced the basic phenomenology of this game and made several new predictions. 

We close by mentioning pertinent open problems. The main problem is that direct experiments are lacking on the validity of Weber's law in the perception of utility changes. It is despite the close attention already paid to Weber's law in economics \cite{luce_jnd,gilboa2009theory,gilboa_cobb,gilboa_interpersonal_utility,monroe1973buyers,kamen1970psychophysics,white2014psychophysics,britt1975weber}. As illustrated by our analysis of the ultimatum game, Weber's law can be used to study two-player bargaining, and more generally partially cooperative behavior. Hence we hope that our results will motivate direct experiments for checking Weber's law in economic experiments. Another interesting open problem is multi-player bargaining setups, where Weber's law can be employed to find new solutions. Also, such setups can be instrumental in suggesting generalizations of Weber's law.  

\subsection*{Acknowledgements}

This work was supported by the Higher Education and Science Committee of Armenia, grant No. 21AG-1C038. 
We acknowledge useful discussions with Lida Aleksanyan and Sanasar Babajanyan.

\bibliographystyle{IEEEtran} 
\bibliography{WF_Bargaining}

\appendix

\section{An alternative implementation of Weber's law in bargaining}
\label{appendix}

Combining Weber's law with bargaining ideas led to several implementation possibilities. In the main text, we presented what we consider to be the most transparent and useful implementation. Alternative implementations, however, may have their own merits, especially in presenting alternative ideas and motivating further thinking. 

Recall discussions around (\ref{simba}--\ref{simba3}) and {\it Axiom 2}, and assume that $\1$ chooses a joint baseline utility $u_1=h[u_2]$ for $\1$ and $\2$. The interests of $\2$ are accounted for by the fact that $\1$ can make yet another step $u_1\to u_1+w_1$ in improving her situation. Eventually, $u_1=u_1^*$ is chosen as follows [we assume $d_1=d_2=0$ for simplicity]:
\BEA
u_1^*={\rm argmax}_{u_1,w_1}\Big[u_1\Big| \frac{h[u_1]-h[u_1+w_1]}{h[u_1]}\leq k_2,\quad \frac{w_1}{u_1}\geq k_1\Big],
\label{a1}
\EEA
where the first inequality in (\ref{a1}) means that $u_1+w_1$ is in the acceptable domain for $\2$, while the second inequality means that $w_1$ will essentially improve the situation for $\1$. A similar strategy can be assumed for $\2$:
\BEA
u_2^*={\rm argmax}_{u_2,w_2}\Big[u_2\Big| \frac{g[u_2]-g[u_2+w_2]}{g[u_2]}\leq k_1,\quad \frac{w_2}{u_2}\geq k_2\Big].
\label{a2}
\EEA
As compared to the {\it Axiom 2} of the main text, Eqs.~(\ref{a1}, \ref{a2}) imply a different type of compromise, and a different type of relevance. A somewhat unnatural point of (\ref{a1}, \ref{a2}) is that the player is given the right to assume baselines for both players. Another point of (\ref{a1}, \ref{a2}) is that $u_1^*$ and $u_2^*$ are not conflicting proposals because working out these inequalities shows that $g[u_2^*]>u_1^*$ and $h[u_1^*]>u_2^*$. This means that there is a certain self-consistency in (\ref{a1}, \ref{a2}).

\end{document}